\documentstyle[aps,prb]{revtex}
\begin{document}
\draft
\twocolumn[\hsize\textwidth\columnwidth\hsize\csname @twocolumnfalse\endcsname

\title{Comment on "$T$- dependence of the magnetic
penetration depth in unconventional superconductors at low
temperatures: Can it be linear?" }
\author{G.E. Volovik }
\address{ Helsinki University of Technology,
Low Temperature Laboratory,
P.O.Box 2200,
FIN-02015 HUT,  Finland, \\
 Landau Institute for Theoretical Physics,
117334 Moscow,
Russia }

\date{\today}
\maketitle

\

]
\narrowtext

In a recent Letter Schopohl and Dolgov (SD) suggested that a
pure $d_{x^2-y^2}$-pairing state becomes invalid in the
zero temperature limit,
$T\rightarrow 0$.\cite{SchopohlDolgov} Their arguments are
based on thermodynamics: if the magnetic
penetration length depends linearly on $T$ at low $T$, the
Nernst theorem -- the third law of thermodynamics -- is
violated. We show here that this conclusion is the result
of the incorrect procedure of imposing the limit
$T\rightarrow 0$ in the electromagnetic response. To
illustrate their reasoning let us consider a simplified case
of the uncharged  Fermi superfluid with lines of zeroes in
the quasiparticle spectrum, the
$d_{x^2-y^2}$-pairing being an example.  In superfluids the
density of the superfluid component $\rho_s(T)$ corresponds
to the magnetic penetration length in superconductors,
$1/\lambda^2(T) \propto \rho_s(T)$. In the case of the
nodal lines it has linear dependence on $T$ at low $T
\ll T_c$:
$\rho_s(T)=\rho - \rho_n(T)$, where the normal component
density in such liquid is
$\rho_n(T)\propto \rho T/T_c$.  The kinetic energy
contribution to the free energy of the liquid flowing with
the superfluid velocity
${\bf v}_s$ along the channel is
\begin{equation}
{\cal F} = {1\over 2} ~\rho_s(T){
v}_s^2 ~,
\label{1}
\end{equation}
We consider the superflow circulating in an annular
channel. This circulation is fixed, if one discards
the negligibly small decay of the
supercurrent via  vortex formation, so one can consider
${\bf v}_s$ as temperature independent. This
results in the  finite  entropy in $T=0$ limit:
\begin{equation}
S(T=0) = - {\partial {\cal F}\over \partial T}\bigg |_{T=0}=
{1\over 2} {\partial
 \rho_n \over \partial T}\bigg |_{T=0}   v_s^2 ~\propto  ~ {
v}_s^2 ~{\rho  \over T_c} ~.
\label{2}
\end{equation}
If one follows the argumentation in
Ref.\cite{SchopohlDolgov}, such a violation of the Nernst
theorem suggests that the superfluid density
$\rho_s$ (or the related penetration length $\lambda$ in
superconductors) cannot be a linear function of
$T$, which would mean that the pairing states with nodal
lines are prohibited at $T=0$ by the Nernst theorem.

There is however a loophole in this argumentation.
The superfluid density $\rho_s(T)$ is the linear response
function of the current ${\bf j}$ to the superfluid velocity
${\bf v}_s$, and thus is obtained in the limit
${\bf v}_s\rightarrow 0$. On the other hand the Nernst
theorem requires the limit
$T\rightarrow 0$ at finite ${\bf v}_s$. These two limits
are not commuting for the kinetic energy ${\cal F}$. The
crossover parameter, $x=T/p_F v_s$, regulates the
scaling behavior of ${\cal F}$ in different limiting cases:
${\cal F}(T,x)  = f(x)\rho v_s^2T/T_c$, where $f(x)$ is
dimensionless function of $x$.\cite{FermionicEntropy} The
regime
$x\gg 1$ corresponds to the linear response to
the superfluid velocity, i.e. to the order of limits when
$v_s\rightarrow 0$ first. In this `high temperature' case,
$T\gg p_Fv_s$, the scaling function
$f(x)\rightarrow {\rm Const}$ and one obtains the finite
entropy, $S(T) = \lim_{T
\rightarrow 0}\lim_{v_s \rightarrow 0}-d{\cal F}/ dT
\propto  \rho v_s^2 / T_c$  in Eq.~(\ref{2}).

In the opposite limit of low
$T$, $x\ll 1$,
the scaling function has the   asymptote
$f(x)\rightarrow  {a\over x} + bx$, where
$a$ and
$b$ are   parameters of order unity.
\cite{FermionicEntropy} In this
true Nernst limit  the entropy is zero at $T=0$:
\begin{equation}
\lim_{v_s
\rightarrow 0}\lim_{T \rightarrow 0}- {d {\cal F} \over dT}
\propto  ~ v_s T ~{\rho  \over p_F T_c} ~,
\label{4}
\end{equation}
in complete agreement with the Nernst theorem. Thus
the linear $T$-dependence of the linear response function
$\rho_s(T)$ does not violate the  third law of
thermodynamics: the Nernst principle does not prohibit a
pure  $d_{x^2-y^2}$-pairing state to exist at
$T=0$ in uncharged Fermi liquid.

The same can be immediately applied to the charged case,
where the superfluid velocity ${\bf v}_s$ is to be
substituted by the external electric current ${\bf j}$
discussed by SD.
\cite{SchopohlDolgov}  Considering the true $T=0$ limit of
the energy,
$\lim_{j
\rightarrow 0}\lim_{T \rightarrow 0} - d{\cal F}/dT
\propto  j T $, one satisfies the Nernst principle. This
does not contradict to the linear
$T$-dependence of the linear electromagnetic response, which
for the wave vector $k=0$ gives
\begin{equation}
\lim_{T
\rightarrow 0} \lim_{j \rightarrow 0}
{d\lambda(k=0,T)\over dT}  =~ {\rm  Const}~.
\label{5}
\end{equation}
For $k\neq 0$  there is another scaling
parameter, $y=T/v_F k$, which regulates  the dependence of
the electromagnetic response on the wave vector $k$
 and produces the $T^2$ dependence
of the penetration length  at
finite $k$, i.e. at $y\ll 1$. \cite{KosztinLeggett} In
opposite case,
$y\gg 1$,  the Eq.~(\ref{5}) is restored.

In conclusion, lines of nodes
in clean superconductors are not in conflict with Nernst
theorem. The answer to the question in the title
of their paper
\cite{SchopohlDolgov} is yes.

\end{document}